\documentclass[aps,prb,reprint,twocolumn,footinbib]{revtex4-2}

\usepackage{graphicx}
\usepackage{bm}
\usepackage{hyperref}
\usepackage{amsmath}
\usepackage{amssymb}
\usepackage{mathtools}
\usepackage{siunitx}
\usepackage{multirow}
\usepackage{makecell}
\usepackage[normalem]{ulem}

\makeatletter
\g@addto@macro\bfseries{\boldmath}
\makeatother

\usepackage{comment}
\usepackage[table,dvipsnames,svgnames]{xcolor}
\usepackage{array}
\usepackage{multirow}
\usepackage[caption=false]{subfig}
\usepackage{braket}
\usepackage{slashed}
\usepackage[english]{babel}
\usepackage[autostyle, english = american]{csquotes}

\usepackage[scr=kp]{mathalpha}

\makeatletter
\def\maketitle{
\@author@finish
\title@column\titleblock@produce
\suppressfloats[t]}
\makeatother

\begin{document}

\title{Composite boson theory of Hall crystals and their transitions to Wigner crystals }
\author{Julian May-Mann} 
\affiliation{Department of Physics, Stanford University, Stanford, CA 94305, USA}
\author{Sayak Bhattacharjee} 
\affiliation{Department of Physics, Stanford University, Stanford, CA 94305, USA}
\author{Srinivas Raghu} 
\affiliation{Department of Physics, Stanford University, Stanford, CA 94305, USA}

\begin{abstract}
{We consider the crystallization of a two-dimensional electron system in a perpendicular magnetic field using composite boson theory. There are three possible states to consider: the Hall liquid, the Wigner crystal, and the Hall crystal (a state with both broken translation symmetry and a quantized Hall response). Within composite boson theory, these states map onto a superconductor, a Mott insulator, and a supersolid of composite bosons respectively. We show that when a $\nu = 1$ Hall liquid has a sufficiently soft roton, there is a first order transition to a triangular lattice Hall crystal. If we continue to decrease the roton mass, there is a continuous transition from the Hall crystal to a Wigner crystal. {When the Hall crystal exhibits the integer quantum Hall effect,} this transition {is} described by a free Dirac fermion and, at the critical point, the coupling to the phonons of the crystal is irrelevant, {in the {renormalization group} sense}. We extend this analysis to fractional $\nu = 1/m$ Hall liquids. There, due to kinetic frustration arising from flux attachment, honeycomb lattice Hall crystals are preferred over triangular ones at intermediate interaction strength. 
}
\end{abstract}

\maketitle
\section{Introduction}
\begin{figure}
    \centering
    \includegraphics[width=0.9\linewidth]{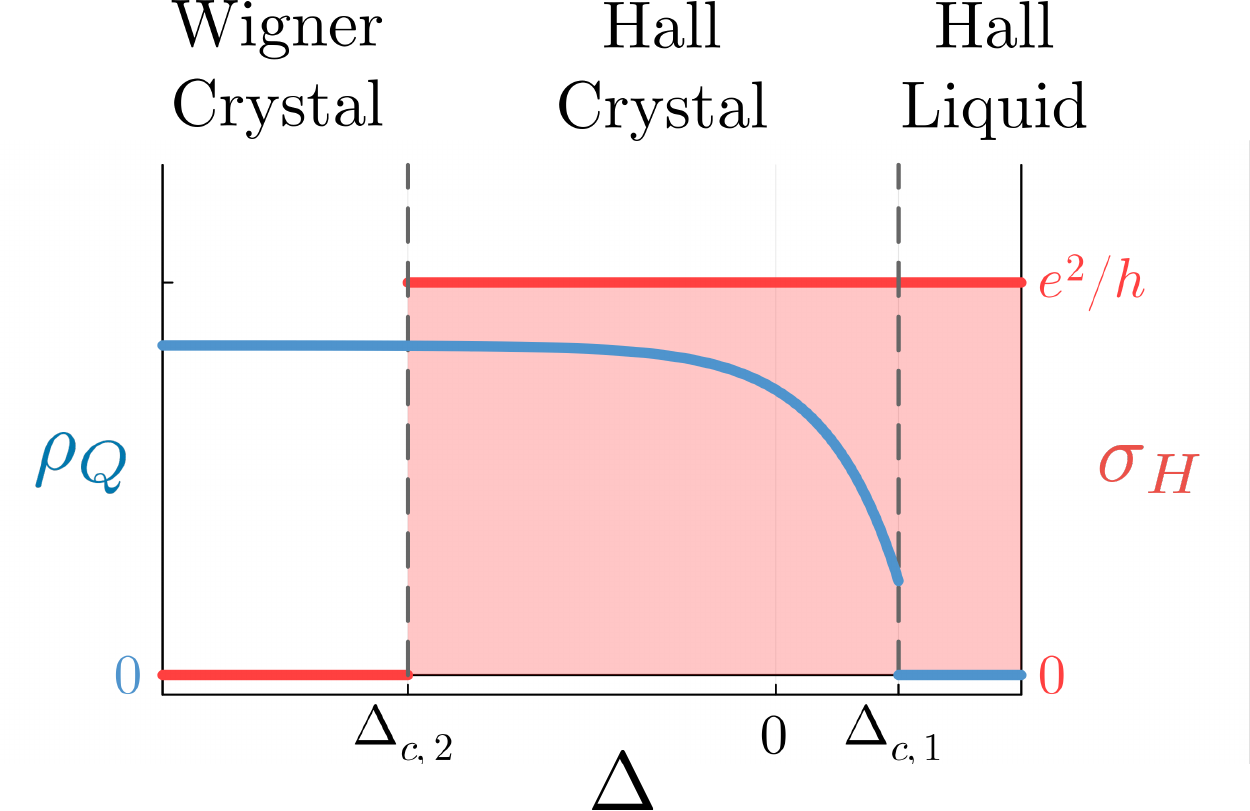}
    \caption{Schematic phase diagram of the composite boson mean-field theory. As a function of the extrapolated roton mass, $\Delta$, there is a first order transition at finite $\Delta_{c,1}>0$ from a Hall liquid to a Hall crystal, followed by a continuous transition to a Wigner crystal at $\Delta_{c,2}<0$. The blue line shows $\rho_Q$, the magnitude of density modulations at finite wavenumber, and the red curve shows  $\sigma_H$, the Hall conductance.}
    \label{fig:placeholder}
\end{figure}
The recent observation of quantized Hall effects in zero magnetic field in {twisted bilayer MoTe$_2$}~\cite{cai2023signatures, park2023observation} and {rhombohedral pentalayer graphene}~\cite{lu2024fractional} have reinvigorated the study of phases of matter with non-trivial topology. Central to this have been questions surrounding the interplay between topology and translation symmetry breaking orders. It was shown in Ref.~\cite{lu2024fractional} that stabilizing an anomalous Hall state requires a weak moir\'e superlattice potential, but if that potential becomes too strong, the state is lost. This led to the conjecture that the anomalous Hall state is actually a weakly pinned “anomalous Hall crystal”~\cite{dong2024anomalous,soejima2024anomalous,tan2024parent,sheng2024quantum,dong2024stability}.

The anomalous Hall crystal is a zero field counterpart of a “Hall crystal” – a phase with both quantized Hall conductance and spontaneously broken {translational symmetry} in a two dimensional electron gas {(2DEG)} subjected to a strong perpendicular magnetic field~\cite{kivelson1986cooperative, halperin1986compatibility, kivelson1987cooperative, Tesanovic, balents1996spatially, murthy2000hall, paul2025designing}. The interplay of translational symmetry breaking and the quantum Hall effect lead to three generic phases: a Hall liquid with quantized Hall conductivity and translation symmetry, a Hall crystal, and a Wigner crystal, where translation symmetry remains broken, but the quantized Hall conductivity is lost. The study of Hall crystals, and their inclusion in the global phase diagram of a 2DEG was initiated in Ref.~\cite{Tesanovic}. Using the Hartree-Fock approximation on a Hamiltonian with specially chosen interactions, the authors of Ref.~\cite{Tesanovic} obtained a phase diagram consisting of Hall liquids, Hall crystals, and Wigner crystals.  
    
In the present work, we analyze the Hall liquid~\cite{klitzing1980new, laughlin1981quantized}, Hall crystals and Wigner crystal~\cite{fukuyama1978pinning} phases of matter, as well as the phase transitions between these phases using the effective field theoretical approach of composite bosons~\cite{zhang1989effective,wen1995topological, fujita2017composite,goldman2020non}. These bosons can be viewed as electrons bound to an odd number of flux quanta. Such an approach is naturally suited to describe the interplay of non-trivial topology associated with the quantum Hall effect and translational symmetry breaking. The composite boson description of the quantum Hall effect makes deep connections between the physics of the quantum Hall effect and superconductivity~\cite{zhang1992chern}. The incompressibility of the Hall liquid can be understood as the Meissner effect of the composite boson superconductor. Extending this further, the Hall crystal phase naturally emerges in this framework as the supersolid order~\cite{leggett1970can,prokof2005supersolid,galli2008solid} of composite bosons, and can further motivate the Wigner crystal as a Mott insulator of composite bosons~\cite{balents1996spatially}. The effective approach presented below has the further advantage of enabling the extension to Hall crystals exhibiting the fractional quantum Hall effect~\cite{, tsui1982two, laughlin1983anomalous}, as well as elucidating the role of elasticity in the neighborhood of a continuous Hall crystal to Wigner crystal transition.  
	
The paper is organized as follows. In Sec~\ref{Sec:Phases}, we overview the Hall liquid, Hall crystal, and Wigner crystal phases. In Sec.~\ref{Sec:CompositeBosons}, we discuss how these phases are manifest in composite boson theory. We discuss the transitions between them in Sec.~\ref{Sec:Transitions}. In Sec.~\ref{Sec:MeanField} we perform mean-field analysis of a filling $\nu = 1$ composite boson model with a soft roton mode. We extend this analysis to fractional $\nu $ states in Sec.~\ref{sec:FracExtension}. We conclude and discuss our results in Sec.~\ref{sec:conclusion}.

\section{Hall Liquid, Hall Crystal and Wigner Crystal}\label{Sec:Phases}
In this section, we briefly review the key distinctions between the Hall liquid, Hall crystal and Wigner crystal.  In a 2DEG with a strong perpendicular magnetic field, $B$, and in the absence of disorder, there is a continuous translational invariance. The Hall liquid preserves translational symmetry and has a quantized Hall conductance. It exhibits a gap to current carrying excitations and the electron density is a rational fraction of the flux density $eB/hc$. The Hall crystal spontaneously breaks translational symmetry but retains a quantized Hall conductivity. As a crystalline state, it has low energy gapless phonon modes and the density of electrons can also vary with respect to the area of the unit cell. Finally, a Wigner crystal results when the Hall conductivity vanishes altogether, and the electron density is commensurate with the crystal. 

These notions are summarized by the 
generalized Středa formula~\cite{Tesanovic},
\begin{equation}
\rho_0 = \nu \frac{eB}{hc} + \eta A^{-1}_0.
\label{eq:generalizedStreda}\end{equation}
$eB/hc$ is the magnetic flux density and $A_0$ is defined as the area of the elementary unit cell of the crystal. For the Hall liquid, $\nu$ is a non-zero rational number and $A^{-1}_0 = 0$ (as the system does not have a crystal unit cell). In this case Eq.~\ref{eq:generalizedStreda} reduces to the conventional Středa formula. In the Hall crystal, $A^{-1}_0 \neq 0$ as there is a well defined crystal unit cell. However, the electron density of the Hall crystal is solely determined by the magnetic flux density, such that $\nu$ is a rational number and $\eta = 0~$\footnote{There are also partial Hall crystals, where $\nu$ and $\eta$ are both non-zero, but such states are not the focus of this work}. Since the Wigner crystal is topologically trivial, $\nu=0$ and $\eta$ is an integer (typically equal to $1$).

A further crucial difference between these phases occurs in their respective spectrum of collective excitations.  In the Hall liquid, all collective excitations, including the long wavelength density fluctuations and the short wavelength rotons, are gapped neutral excitations\cite{kallin1984excitations, girvin1985collective}. As the roton minimum gap starts to close, the Hall liquid gives way to a crystal state via roton condensation.  In both  the Hall and  Wigner crystals, gapless collective excitations occur at long wavelength and are the Nambu-Goldstone modes associated with translational symmetry breaking~\cite{dong2025phonons, hirsbrunner2026topological}, i.e.  acoustic phonons.  

In the subsequent sections, we show  that the salient features of the Hall liquid, Hall crystal and Wigner crystal are manifest within composite boson theory that was originally developed to describe Hall liquids~\cite{zhang1989effective}.

\section{Composite Bosons}\label{Sec:CompositeBosons}
The idea of ``flux attachment" and statistical transmutation is useful in theoretical studies of 2DEG at high magnetic fields~\cite{zhang1989effective, fradkin1989jordan, jain2007composite}. 
For instance, a physical electron can be viewed as a bound state of a ``composite'' boson and an odd number of flux quanta of an emergent $U(1)$ gauge field. 
When there are $m$ external magnetic flux quanta per electron and $m$ is odd (i.e. a $1/m$ filled Landau level), the composite bosons experience no net flux on average. This enables a study of electrons in a magnetic field using bosons in zero magnetic field. 

The composite boson Lagrangian for a system of electrons at filling $\nu = 1/m$ can be written as (using $e = \hbar = c = 1$),
\begin{equation}\begin{split}
&\mathcal{L} = \mathcal{L}_{\text{B}} + \mathcal{L}_{\text{CS}},\\
&\mathcal{L}_{\text{B}} = \Phi^* \left(i\partial_t + a_t - A_t\right)\Phi\\&\phantom{===}- \frac{1}{2M^*} |\left(i \bm{\nabla} + \bm{a} - \bm{A}  \right) \Phi|^2 + \mathcal{L}_{\text{int}},\\
&\mathcal{L}_{\text{CS}} = \frac{\epsilon^{\mu\nu\lambda}}{4\pi m} a_\mu \partial_\nu a_\lambda.
\label{eq:GenLagrangian}\end{split}\end{equation}
where $\Phi$ is a complex boson with effective mass $M^*$, $a_\mu$ is the emergent gauge field, and $A_\mu$ is the physical background gauge field. Here, $A_\mu$  is a non-dynamical gauge field and satisfies $\bm{\nabla}\times \bm{A} = 2\pi m \rho_0$, where $\rho_0$ is the average density of bosons (also equal to the density of the underlying electrons). In the following, the interaction term $\mathcal{L}_{\text{int}}$ will be a density-density interaction between the bosons, but we will leave it arbitrary at present.

The Chern-Simons term in the Lagrangian, $\mathcal{L}_{\text{CS}}$, attaches $m$ fluxes of $a_\mu$ to each boson. Using the equations of motion for $a_t$, we obtain the following ``flux attachment'' constraint,
\begin{equation}\begin{split}
\bm{\nabla}\times \bm{a} &= -2\pi m |\Phi|^2. 
\label{eq:fluxAttachEqs}\end{split}\end{equation}
We note that the Chern-Simons term in Eq.~\ref{eq:GenLagrangian} is not properly quantized when defined on a closed manifold. However, this issue is easily resolved by introducing auxiliary gauge fields, and will not be important for our analysis\cite{seiberg2016duality}. 

\begin{table}
\centering
\renewcommand{\arraystretch}{2} 
\setlength{\tabcolsep}{5pt}      
\begin{tabular}{c|c|c|c}
 & \makecell{Hall\\ Liquid} &  \makecell{Hall\\ Crystal} & \makecell{Wigner\\ Crystal} \\ \hline\hline
\makecell{$\rho(\bm{r})$ \\ (local density)} & Constant & \makecell{Modulated} & \makecell{Modulated} \\\hline
\makecell{$a_\mu$, $A_\mu$ \\ (gauge fields)} & Higgsed & Higgsed & Not Higgsed\\ \hline
\end{tabular}
\caption{Summary of the salient features of the Hall liquid, Hall crystal and Wigner crystal in composite boson theory. }
\label{tab:PhaseSummary}
\end{table}

In this work, we are primarily interested in $\nu = 1$, where the electrons are at integer filling of the Landau level. Nevertheless, as the only difference between fractional and integer states in composite boson theory is the coefficient of the Chern-Simons term, it is straightforward to generalize the construction we use for $\nu = 1$ to $\nu  = 1/m$. This will be done in Sec.~\ref{sec:FracExtension}.

The rest of this section is devoted to discussing how the phases of interest (Hall liquid, Hall crystal and Wigner crystal) are realized in composite boson theory at $\nu= 1$. As we shall discuss, these phases are distinguished by (1) whether the density modulates, and (2) whether the gauge fields are massive due to the Higgs mechanism induced by composite boson condensation. This is summarized in Table~\ref{tab:PhaseSummary}. 

The Hall liquid 
corresponds to the uniform superfluid/superconductor of composite bosons~\cite{zhang1989effective}. In the Lagrangian formulation (Eq.~\ref{eq:GenLagrangian}), this is the saddle point where $\Phi = \sqrt{\rho_0}$ and $a_\mu = A_\mu$. To confirm that this state has a quantized Hall response, we add a probe component to the background gauge field $A_\mu \rightarrow A_\mu + \tilde{A}_\mu$, and consider fluctuations around the liquid saddle point. The emergent gauge field $a_\mu$ is Higgsed in this phase, and so the matter and gauge fluctuations around the liquid state are massive. After integrating over the massive fluctuations, the effective response theory for $\tilde{A}_\mu$ is simply the Chern-Simons term,
\begin{equation}
    \mathcal{L}_{\text{response}} = \frac{\epsilon^{\mu\nu\lambda}}{4\pi} \tilde{A}_\mu \phantom{|}\partial_\nu \phantom{|} \tilde{A}_\lambda,
\end{equation}
and we find $\sigma_H = 1/2\pi$, as expected. 

We now turn to the composite boson description of the Hall crystal, which will be a central focus of this work. Since the Hall liquid is a superfluid of composite bosons, the Hall crystal naturally corresponds to a \textit{supersolid} of composite bosons. A conventional supersolid is a state that breaks both translational symmetry and the $U(1)$ symmetry of the bosons, leading to coexisting superfluidity and crystalline order. For composite bosons, this will be a state where the local boson density $\rho(\bm{r})$ periodically modulates, and the gauge fields remain Higgsed. If we imagine a pinned lattice (i.e. no phonons) the gauge and matter fluctuations around the Hall crystal state will both be massive, and by the same logic used above, $\sigma_H = 1/2\pi$. Without pinning one has to account for the sliding motion of the crystal, and transport is likely no longer quantized.

Finally, there is the composite boson description of the Wigner crystal. Like the Hall crystal, this will be a state where the local boson density periodically modulates. However, for the Wigner crystal, the gauge fields are not Higgsed. Without the Higgs mechanism, there is no energy cost for adding or removing fluxes of the background gauge field. The composite boson density is therefore not tied to the magnetic field, and the electron density no longer follows Eq.~\ref{eq:fluxAttachEqs}. The density is instead fixed by the structure of the crystal lattice. Note that the average of the local boson density, $|\Phi|^2$, remains finite for the Wigner crystal even though the gauge fields are not Higgsed.

\section{Phase Transitions}\label{Sec:Transitions}
Having discussed the different phases in composite boson theory, we now analyze the transitions between them. The Hall liquid to Wigner crystal transition is known to be first order~\cite{lam1984liquid}. As this transition has been well studied in other works, we will focus our discussion on the Hall liquid to Hall crystal and Hall crystal to Wigner crystal transitions. Our discussion of these transitions follows from our previous discussion, and will be verified in our mean-field analysis in Sec.~\ref{Sec:MeanField}.

\subsection{Hall liquid to Hall crystal transitions}
The transition between the Hall liquid and the Hall crystal is a conventional crystallization transition. The topological characteristics of the two phases are the same. The two phases are therefore distinguished by a local order parameter, and the transition can be described within Landau theory. Based on this, the crystallization transition is expected to be first order, and to lead to a triangular lattice crystal. Both of these features can be attributed to a tri-linear term that is present in the expansion of the triangular lattice free energy, but is absent for other crystals (see Appendix~\ref{app:FreeEnergyTrianglular}). 

\subsection{Hall crystal to Wigner crystal transition}
The transition between the Hall crystal and Wigner crystal has a more interesting structure. This is a topological transition where the Hall conductance of the system changes. In the composite boson framework, the Wigner to Hall crystal transition corresponds to the Higgsing of the gauge fields $a_\mu$ and $A_\mu$. The crystal lattice is not expected to change during this transition. For our initial discussion we will also ignore phonons, and take the crystal lattice to be fixed. As we shall show, the coupling to phonons is irrelevant (at $\nu=1$), and their presence does not change the details of the transition. 

If we fix the crystal lattice, the transition between the Hall crystal and Wigner crystal has a striking similarity to the transition between a superconductor to Mott insulator on a lattice. In the Hall crystal/superconductor, the gauge fields are Higgsed, and in the Wigner crystal/Mott insulator, they are not. The only difference comes from the Chern-Simons term. The superconductor to Mott insulator transition is well studied, and is known to be continuous~\cite{fisher1989boson}. Physically, the transition occurs when the vortices of the superconductor proliferate, leading to the Mott insulating phase where the Cooper pairs lose phase coherence. 

Based on this, the Hall crystal to Wigner crystal transition will involve the proliferation of vortices of $\Phi$. The Lagrangian for this transition is,
\begin{equation}\begin{split}
\mathcal{L}_{\textrm{vort}} = &\frac{1}{2} |D^\alpha_\mu \phi_v|^2 - r_v|\phi_v|^2 - u |\phi_v|^4 \\ &- \frac{1}{4\pi} \alpha d \alpha + \frac{1}{2\pi}  \alpha d\tilde{A}. 
\label{eq:VortexLag}\end{split}\end{equation}
where  $D^\alpha_\mu \equiv \partial_\mu - i \alpha $, and  $ada \equiv \epsilon^{\mu\nu\lambda}\alpha_\mu\partial_\nu \alpha_\lambda$. The coarse grained vortex fluctuations are encoded in $\phi_v$, and $\alpha$ is a new emergent gauge field. We review the argument for arriving at this Lagrangian through the boson-vortex duality~\cite{lee1991anyon, peskin1978mandelstam, zhang1992chern} starting from Eq.~\ref{eq:GenLagrangian} in Appendix~\ref{vortex_theory}.

The Hall crystal and the Wigner crystal correspond to phases with  confined and proliferated vortices respectively. For $r_v>0$ the vortices are confined and decouple from the the gauge field $\alpha$. If we integrate out both $\phi_v$ and $\alpha$, the effective response theory is the usual Chern-Simons term for $A$, as expected for the Hall crystal. For $r_v<0$, the gauge field $\alpha$ acquires a Higgs mass, and, after integrating out $\phi_v$ and $\alpha$, there is no Chern-Simons term for $A$ as expected for the Wigner crystal.  Furthermore, when $r_v<0$, it is possible to change the magnetic field without changing the physical density. This follows from the equations of motion of $\alpha_t$,
\begin{equation}\begin{split}
\frac{1}{2\pi} \bm{\nabla}\! \times\! \left[\tilde{\bm{A}} - \bm{\alpha}\right] = \frac{1}{2} \left[ i \phi_v \partial_t \phi_v^* \!- i \phi^*_v \partial_t \phi_v\right] +  \alpha_t |\phi_v|^2.
\end{split}\end{equation}
We can see that an external magnetic field can be screened by increasing the vortex density (i.e. adding a time dependent phase to $\phi_v$). Importantly, this can be done without changing the physical density,  $\frac{1}{2\pi} \bm{\nabla} \times \bm{\alpha}$. The density is therefore independent of the background magnetic field. Following our discussion of Eq.~\ref{eq:generalizedStreda} confirming that this state is a Wigner crystal. 

It is worth pointing out that the vortex theory has dynamical critical exponent $z = 1$, despite the fact that there is a $i\Phi^* \partial_t \Phi$ term in Eq.~\ref{eq:GenLagrangian}. This is because there is an emergent particle-hole/vortex-antivortex symmetry at $\nu =1$ that forbids a linear time-derivative term for $\phi_v$. To demonstrate why this is, let us imagine that the theory is instead $z = 2$, so that there is a term $\propto i\phi_v^* \partial_t \phi_v$. If this is the case, the vortex density would be $\propto |\phi_v|^2 \geq 0$. However, since both positive and negative vorticity solutions are possible, the vortex density cannot be positive semi-definite, and the theory must therefore have $z=1 $. 

An advantage of the effective field theory approach is that it enables us to understand the effect of additional gapless degrees of freedom on the Hall to Wigner crystal transition.  Natural candidates for additional gapless degrees of freedom are the elastic modes associated with phonons of the crystal.  To include phonons in the critical theory, we use a boson-vortex  duality, to convert the vortex theory in Eq.~\ref{eq:VortexLag} to a theory for the matter fields, $\phi$,
\begin{equation}\begin{split}
\mathcal{L}_{\text{mat}} = &\frac{1}{2} |D^{b-\tilde{A}}_\mu \phi|^2 - r_m|\phi|^2 - u' |\phi|^4 - \frac{1}{4\pi} bdb, 
\label{eq:MatterLag}\end{split}\end{equation}
where $b$ is a new emergent gauge field. Again, this is a $z = 1$ theory. In the matter theory, $r_m<0$ is the topologically non-trivial Hall crystal and $r_m>0$ is the topologically trivial Wigner crystal phase, as is clear from integrating out $\phi$ and $a$ in either phase.

Phonons are included via the elastic deformation vector $\bm{u}$, and linearized strain tensor $u_{ab} = (\partial_a u_b + \partial_a u_b)/2$, $a, b = x,y$. Including these fields, the critical theory is
\begin{equation}\begin{split}
&\mathcal{L} = \mathcal{L}_{\text{mat}}+\mathcal{L}_{\text{ph}}+\mathcal{L}_{\text{mat-ph}},\\ 
&\mathcal{L}_{\text{ph}} =  \frac{\rho_m}{2} |\partial_t \bm{u}|^2 + \lambda_{abcd} u_{ab}u_{cd}, \\ 
&\mathcal{L}_{\text{mat-ph}} =  g_{ab} u_{ab} |\phi|^2,
\label{eq:MatterAndPhLag}\end{split}\end{equation}
where $\rho_m$ is the mass density associated with the elastic deformation, and the couplings $\lambda$ determine the phonon spectrum. The $g$ coupling is the lowest order symmetry preserving coupling between the matter and phonon fields. Since this model has inversion symmetry, a kineo-elastic term is not allowed~\cite{dong2025phonons}. We have, however, not included a symmetry allowed Berry curvature term, $\propto u \partial_t u$ as previous works have shown that this term is zero in the Hall crystal phase~\cite{Tesanovic, soejima2025topological}.

Although phonons are generically present their coupling to matter in Eq.~\ref{eq:MatterAndPhLag} is irrelevant. This can be directly observed by using a boson-fermion duality~\cite{seiberg2016duality, Chen2018, senthil2019duality} to rewrite $\mathcal{L}_{\text{mat}}$ in terms of a Dirac fermion, 
\begin{equation}\begin{split}
\mathcal{L}_{\text{mat}} = i \bar{\Psi} \slashed{D}^{\tilde{A}} \Psi - m_D \bar{\Psi} \Psi + \frac{1}{8\pi} \tilde{A} d  \tilde{A} .
\label{eq:DiracDualized}\end{split}\end{equation}
Since integrating out a massive Dirac fermion leads to a Chern-Simons term with coefficient $\text{sign}(m_D)/8\pi$, we identify the $m_D>0$ phase with the Hall crystal and the $m_D<0$ phase with the Wigner crystal. Under this duality, $|\phi|^2$ maps to $-\bar{\Psi} \Psi$, and, because of this, the matter-phonon coupling takes on the general form
\begin{equation}\begin{split}
\mathcal{L}_{\text{mat-ph}} =  g'_{ij} u_{ij} \bar{\Psi} \Psi.
\label{eq:DiracPhCoup}\end{split}\end{equation}
Since tree level scaling dimension of the quantum fields is $[u_{ij}] = 3/2$, and $[\Psi] = 1$, the phonon coupling is irrelevant by simple power counting. The effects of phonons can therefore be ignored at the Hall crystal to Wigner crystal transition. However, as we shall show in Sec.~\ref{sec:FracExtension}, this conclusion does not hold when considering the transition from a fractional Hall crystal to Wigner crystal.


Based on Ref.~\cite{fisher1989boson}, the critical theory in Eq.~\ref{eq:MatterLag} also describes the critical behavior of Bose-Hubbard model coupled to a Chern-Simons term. The fixed lattice of this model should be equated with the underlying crystalline structure of both the Hall crystal and Wigner crystal phases. Since the lattice structure does not change during the Hall crystal to Wigner crystal transition, and the phonons are irrelevant at the transition, treating the lattice as fixed is a valid assumption. A detailed analysis of such a lattice model has been relegated to a companion work.

\section{Mean Field Analysis}\label{Sec:MeanField}
The low energy effective theory above adequately captures universal aspects of a putative continuous Hall to Wigner crystal transition.  We now show that the same theory also captures the key properties of the ensuing phases themselves.  
To simplify this discussion we will use dimensionless units, where the average composite boson density $\rho_0 = 1$ and the cyclotron frequency is $\omega_c = 2\pi$. In these units, the composite boson Lagrangian is 
\begin{equation}\begin{split}
\mathcal{L}_{\text{B}} = &\Phi^* \left(i\partial_t + a_t-A_t\right)\Phi- \frac{1}{2} |\left(i \bm{\nabla} + \bm{a} -\bm{A} \right) \Phi|^2 \\ &- \frac{1}{2}\int d^2\bm{r}' \phantom{|} (\rho(\bm{r})-1) V(|\bm{r}-\bm{r}'|)(\rho(\bm{r}')-1).\\
\label{eq:rescaledLagrangian}\end{split}\end{equation}
We have taken the interaction, $\mathcal{H}_{\text{int}}$ to be a density-density interaction between the boson, $V(|\bm{r}|)$, which will be treated within mean-field theory.

\begin{figure}[t!]
\centering
\includegraphics[width = .9\linewidth]{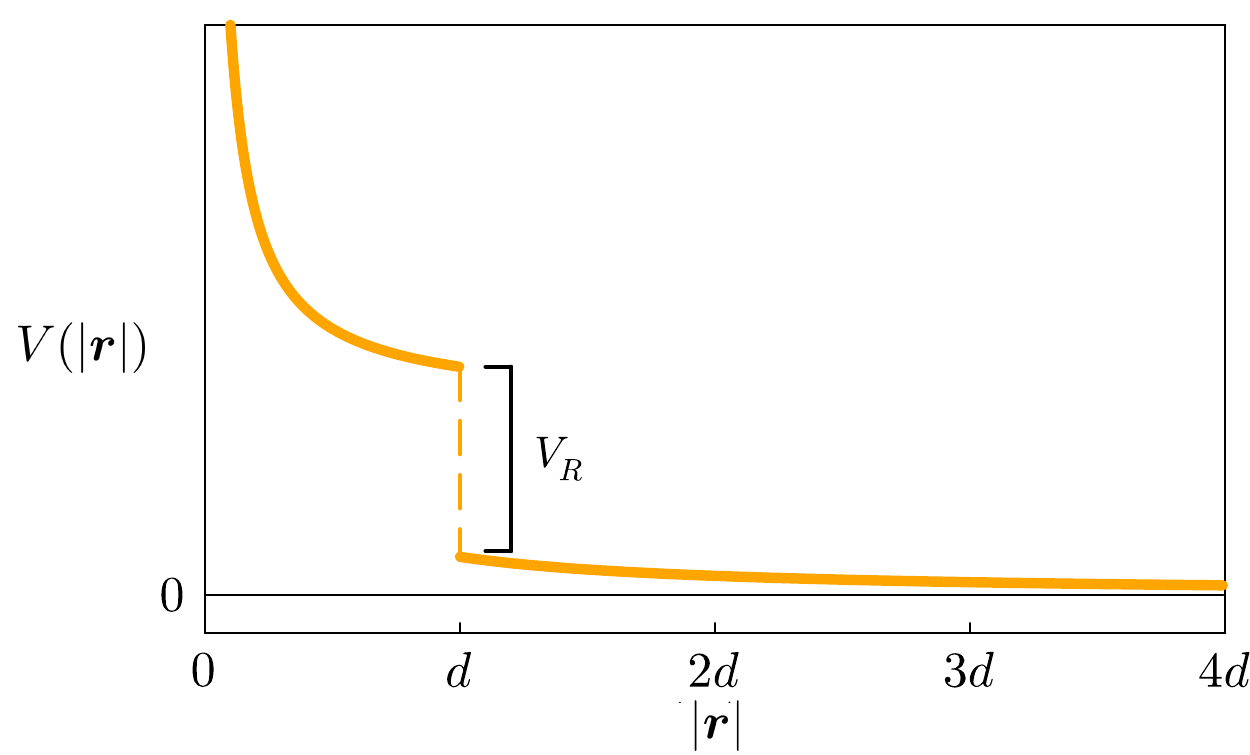}
\caption{Functional form of the real space interaction. At long distances the interaction is the $1/|\bm{r}|$ Coulomb interaction, $V_C$. As short distances there is a discontinuity due to the phenomenological interaction $V_{R}$. }\label{fig:interactionFig}
\end{figure}

For our modeling, we assume that the crystallization arises due to a soft roton mode\cite{girvin1986magneto}. The roton mode arises from many-body effects, and, cannot be captured from mean-field analysis of the Coulomb interaction alone in composite boson theory. To include the roton mode, we will therefore consider an effective interaction that involves both Coulomb repulsion and an additional phenomenological short range repulsive interaction. The interaction term (in dimensionless units) is,
\begin{equation}\begin{split}
V(|\bm{r}|) = V_{C}\frac{1}{|\bm{r}|} + V_R \Theta(d - |\bm{r}|),
\label{eq:intDef}\end{split}\end{equation}
where $V_{C}>0$ is the strength of the Coulomb interaction, $V_R>0$ is the strength of the phenomenological short range repulsive interaction, and $\Theta$ is the Heaviside step function. Note that the effective interaction is repulsive at all distances, as shown in Fig.~\ref{fig:interactionFig}. As we shall show, $V_R$ leads to a soft roton mode within mean-field theory\cite{pomeau1993model, during2011theory}, and should be understood as mimicking the complex many-body effects that are responsible for the roton minimum. 

\subsection{Roton minimum}\label{sub:roton_minimum}
To show that the phenomenological interaction $V_R$ leads to a roton mode, we consider density fluctuations around the liquid state of Eq.~\ref{eq:rescaledLagrangian}. In the original problem of electrons in a magnetic field, these are inter-Landau fluctuations. 

Since the composite boson liquid is, effectively, a superconductor, the spectrum of density fluctuations can be found using Bogoliubov theory~\cite{bogoliubov1947theory}. In practice, this involves solving the linearized equations of motion for the density and gauge field fluctuations around the liquid state, $\Phi = 1$, $a_\mu = A_\mu$. For a density wave $\delta\rho \propto \cos(|\bm{p}| x+\omega t)$ the dispersion is
\begin{equation}\begin{split}
\omega =  \sqrt{4\pi^2+ \frac{|\bm{p}|^4}{4} + 2\pi V_C |\bm{p}|  + 2\pi  V_R d |\bm{p}|   J_1(|\bm{p}|d)   },
\label{eq:interLLSpectrum}\end{split}\end{equation}
where $J_n$ is the $n^\text{th}$ Bessel function. The spectrum has a gap of $\omega_c = 2\pi$ at $\bm{p} = 0$ as required by Kohn's theorem~\cite{kohn1961cyclotron}. Without interactions, the gapped modes have a quadratic dispersion at small momentum. The Coulomb interaction makes the dispersion linear at small $\bm{p}$.

\begin{figure}
\centering
\includegraphics[width = .9\linewidth]{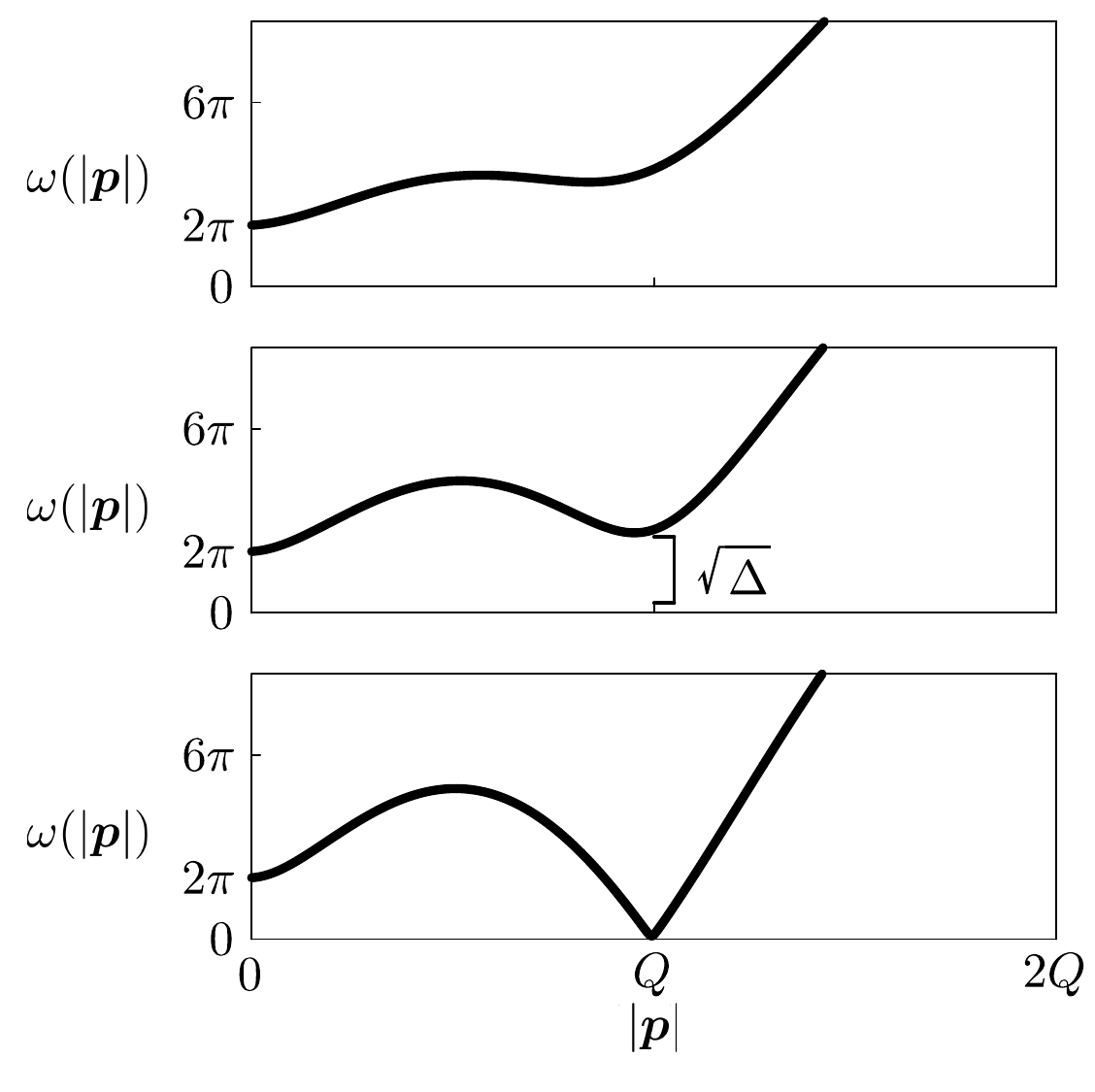}\caption{The dispersion of density fluctuations, $\omega(|\bm{p}|)$ for $V_C = 50$, and $V_R = 8$ (top), $V_R = 8$ (middle), and $V_R\approx 22.05$ (bottom) where the roton gap closes. $\omega(Q) = \sqrt{\Delta}$ is indicated in the middle plot.}\label{fig:roton_Disperision}
\end{figure}

When the phenomenological short ranged interaction, $V_R$ is significantly large, the dispersion in Eq.~\ref{eq:interLLSpectrum} has a finite momentum roton minimum, as shown in Fig.~\ref{fig:roton_Disperision}. This occurs when  
\begin{equation}\begin{split}
\frac{x^3}{d^3} + 2\pi V_C + 2\pi V_R dx J_1(x) = 0
\end{split}\end{equation}
has a non-trivial solution for some $x>0$. If we further increase $V_R$, the roton mode becomes softer, and eventually become gapless at finite momentum. The momentum where this occurs can be controlled via $d$. As we are assuming that the roton mode is a precursor to a triangular lattice Wigner crystal in this modeling, we fix $d$ such that $Q = \pi \sqrt{8/\sqrt{3}}$. In practice this makes $d$ a function of $V_C$ (for example, $d \approx .77$ for $V_C = 50$). 

Since the softening of the roton leads to crystallization, it will be useful to define the extrapolated roton mass
\begin{equation}\begin{split}
\Delta = 4\pi^2+ \frac{Q^{4}}{4} + 2\pi V_C Q  + 2\pi  V_R d Q   J_1(Q d).
\end{split}\end{equation} 
This will be the main theoretical tuning knob for our analysis. For $\Delta > 0$ the roton is gapped, becoming gapless at $\Delta = 0$. For $\Delta < 0$ the system is unstable to forming a crystal state. As noted previously, the liquid to triangular lattice crystal transition is expected to be first order. In the next section, we will show that this is indeed the case here, and that the transition occurs at finite $\Delta> 0$, when the roton is still massive.

\subsection{Crystal ansatz}

\begin{figure}[t!]
\centering
\includegraphics[width = .9\linewidth]{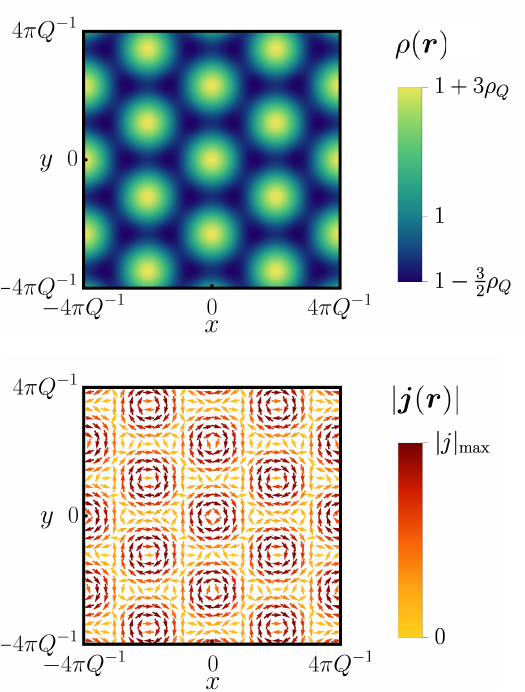}
\caption{Top) The density profile in Eq.~\ref{eq:crystalAnsatzDensity} for a triangular crystal with $\rho_Q > 0$. Bottom) the current profile of the crystal, showing a circulating pattern of currents. }\label{fig:DensityPlotFirst}
\end{figure}
Since we are interested in crystals that arise from a soft roton, we will consider crystals where there is a smooth variation in the density of the composite bosons. We will also assume that $\Phi$ is free of vortices. This is reasonable when the finite energy cost associated with the vortex core is larger than the cyclotron frequency. Under these assumptions, we can fix $\Phi$ to be real, $\Phi = \sqrt{\rho}$, and use the following general crystal ansatz for the density profile,
\begin{equation}\begin{split}
&\rho(\bm{r}) = 1 + \sum_i \rho_{\bm{Q}_i} \cos(\bm{Q}_i \cdot \bm{r}). 
\label{eq:crystalAnsatzDensity}\end{split}\end{equation}
Here, $\rho_{\bm{Q}_i}$ are the variational parameters, and $\bm{Q}_i$ are the ordering vectors of a given crystal. Due to the soft roton mode at $Q$, the energetically favorable crystals have $|\bm{Q}_i| = Q$. We will also assume rotation symmetry of the crystal, such that we can set $\rho_{\bm{Q}_i} = \rho_{Q}$ for all $i$. Due to the flux attachment constraints, the crystal ansatz also induces modulations in the gauge fields. For the density profile in Eq.~\ref{eq:crystalAnsatzDensity}, the gauge field configurations, 
\begin{equation}
\begin{split}
    &a_x = A_x +\frac{2\pi}{ |Q|^2} \partial_y \rho(\bm{r}),\\
    &a_y = A_y -\frac{2\pi}{ |Q|^2} \partial_x \rho(\bm{r}),
\end{split}\label{eq:crystalAnsatzGauge}
\end{equation}
satisfy the flux attachment constraints. 

In additional to conventional crystalline order, this ansatz also has an accompanying pattern of orbital currents resembling 
$d$-density wave order~\cite{Chakravarty2001}, which manifests as a spatially modulated pattern of circulating currents in the ground state,
\begin{equation}\begin{split}
\bm{j} = \bm{j}_p+\bm{a} \rho,
\label{eq:CurrentsForDDWave}\end{split}\end{equation}
where $\bm{j}_p$ is the paramagnetic part of the current, which is zero for our ansatz. The currents are generically non-zero as the spatial fluctuations of $\rho$ necessarily lead to fluctuations of $\bm{a}$ due to the flux attachment constraint, Eq.~\ref{eq:fluxAttachEqs}. The currents circulate around the maxima and minima of the density profile, with opposite chiralities, as can be seen from taking the curl of Eq.~\ref{eq:CurrentsForDDWave} and using the flux attachment constraints. In Fig.~\ref{fig:DensityPlotFirst}, we plot a representative density and current profile for a triangular lattice crystal. 

There are several possible types of translation symmetry breaking orders that one can consider. First, there is the triangular crystal, where the sum in Eq.~\ref{eq:crystalAnsatzDensity} is over $\bm{Q}_1 = (1,0)Q$, $\bm{Q}_2 = (-\frac{1}{2},\frac{\sqrt{3}}{2})Q$, and $\bm{Q}_3 = (-\frac{1}{2},-\frac{\sqrt{3}}{2})Q$. Since $\rho \geq 0$, $\rho_Q$ must be between $-1/3$ and $2/3$ for the triangular lattice. Second, there is the square lattice where the sum is over $\bm{Q}_1 = (1,0)Q$ and  $\bm{Q}_2 = (0,1)Q$. For the square lattice, $\rho_Q$ must be between $\pm 1/2$.  Third, there is stripe order, where the density only involves $\bm{Q}_1 = (1,0)Q$. For the stripe order, $\rho_Q$ must be between $\pm 1$.

\begin{figure}
\includegraphics[width = .9\linewidth]{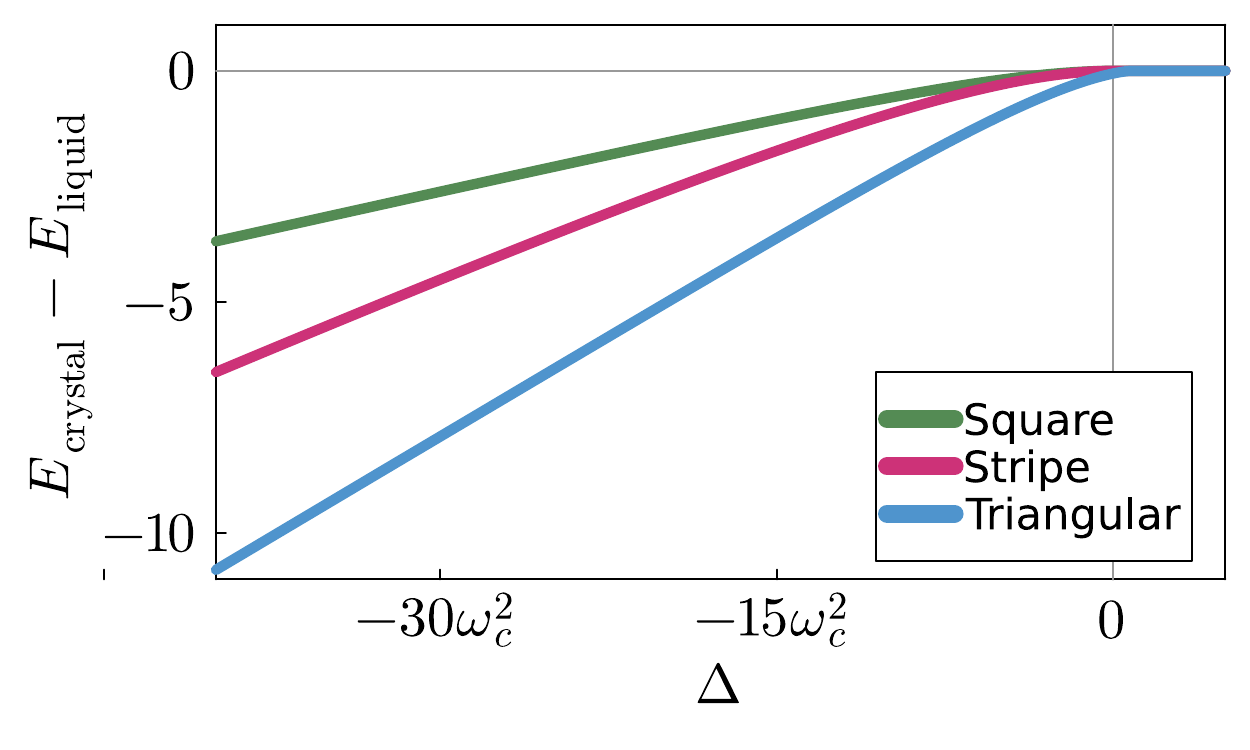}
\caption{The difference in energy density between the different crystal states and the liquid state, as a function of the roton mass $\Delta$. The triangular crystal always has the lowest energy.}\label{fig:CrystalEnergies}
\end{figure}

To determine the dominant translation symmetry breaking order, we calculate the optimal energy for the different ansatze using the Hamiltonian density,
\begin{equation}\begin{split}
\mathcal{H} = &\frac{1}{2} |(i \bm{\nabla}+\bm{a}-\bm{A})\Phi| \\ &+ \int d^2 \bm{r}' (\rho (\bm{r})-1)V(|\bm{r}-\bm{r}'|)(\rho (\bm{r}')-1).
\label{eq:HamFirstHarm}\end{split}\end{equation}
 The optimal energies as a function of $\Delta$ are shown in Fig ~\ref{fig:CrystalEnergies}. The triangular crystal has the lowest energy, and so we will restrict out attention to the triangular ansatz for the remainder of this work. 


\subsection{Mean-field phases}

\begin{figure}[t!]
\centering
\includegraphics[width = .9\linewidth]{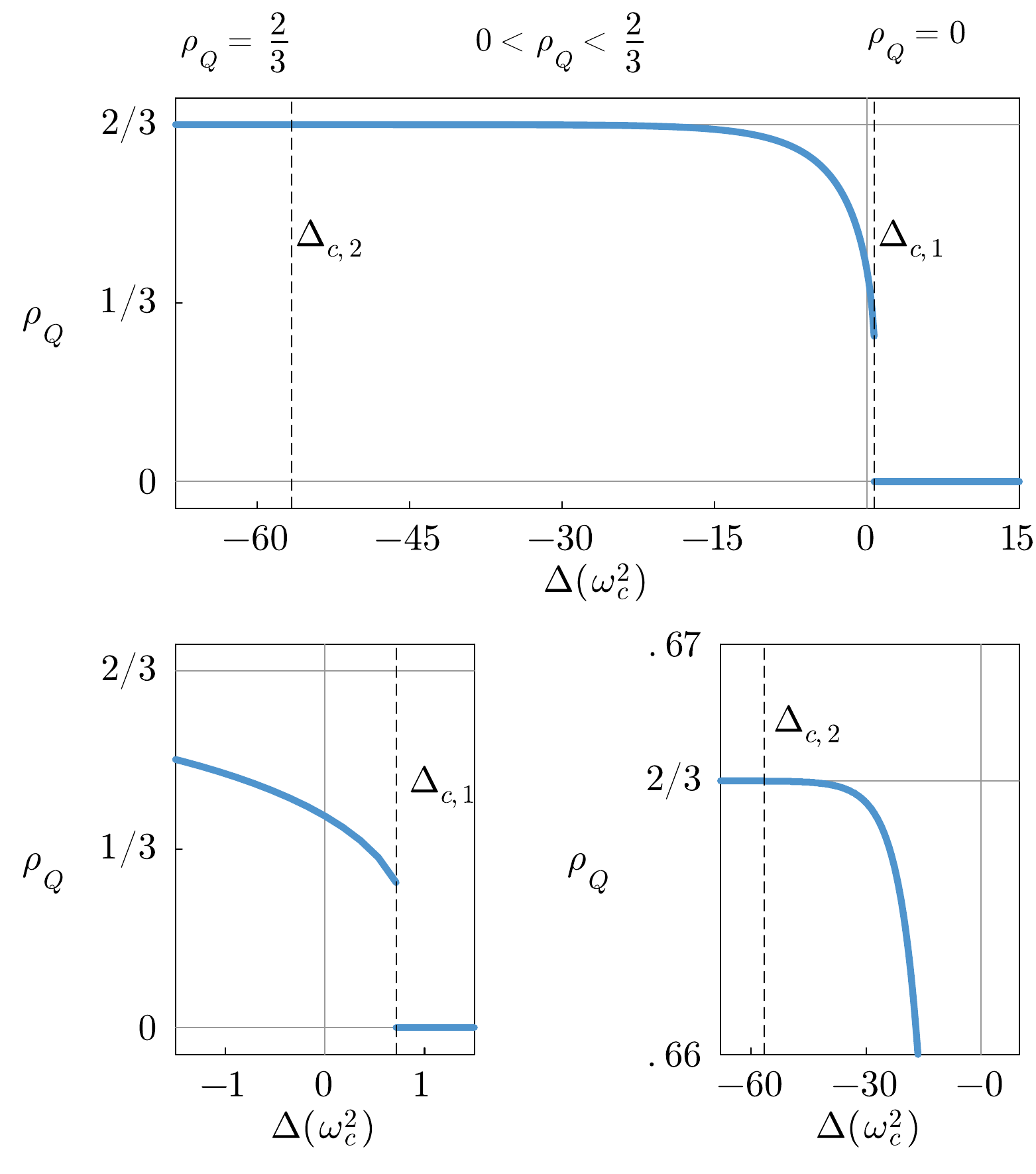}
\caption{Top) The optimal value of the the variational parameter, $\rho_Q$ as a function of the roton mass $\Delta$. There is a first order liquid to crystal transition at $\Delta = \Delta_{c,1} \approx .66 \omega^2_c$ and a continuous topological transition at $\Delta_{c,2}\approx 58 \omega^2_c$. Bottom) Same zoomed in around, $\Delta_{c,1}$ and $\Delta_{c,2}$.  Optimal values were calculated using a $1500\times1500$ point real space grid and sampling $10^6$ possible $\rho_Q$'s.}\label{fig:Variational_Result_1st_Harm}
\end{figure}

For the triangular crystal, the phase diagram is determined from the optimal value of $\rho_Q$ that minimizes Eq.~\ref{eq:HamFirstHarm}. It is worth noting that the positive and negative $\rho_Q$'s are distinct from one another, as the density maxima form a triangular lattice for $\rho_Q > 0$ and a honeycomb lattice for $\rho_Q < 0$. In principle, there are five possible phases to consider: Hall liquid, triangular Hall crystal, triangular Wigner crystal, honeycomb Hall crystal, and a trivial honeycomb crystal with vanishing Hall conductance. As shall be demonstrated in the following subsections: $\rho_Q = 0$ is the Hall liquid, $0<\rho_Q <2/3$ is the triangular Hall crystal, $\rho_Q = 2/3$ is triangular Wigner crystal, $0>\rho_Q > -1/3$ is the honeycomb Hall crystal, and $\rho_Q = -1/3$ is the trivial honeycomb crystal. 

The optimal value of $\rho_Q$ as a function of $\Delta$ is shown in Fig.~\ref{fig:Variational_Result_1st_Harm}. As a function of decreasing $\Delta$, the liquid phase is stable up to $\Delta_{c,1}>0$. At this point there is a first order transition to a triangular Hall crystal, where $\rho_Q$ abruptly becomes positive. As $\Delta$ further decreases, $\rho_Q$ smoothly increases, until finally reaching $\Delta_{c,2}$, where it reaches the critical value of $\rho_Q = 2/3$  and the system enters a triangular Wigner crystal phase. The triangular Wigner crystal phase is stable for all $\Delta < \Delta_{c,2}$. Numerically solving for the mean-field energy, we find that $\Delta_{c,1} \approx .66 \omega_c^2$ and $\Delta_{c,2} \approx - 58 \omega_c^2$

Since the triangular lattice always has lower energy than the honeycomb lattice here, we will only explicitly discuss the triangular Hall crystal and triangular Wigner crystal in the following (which we shall simply refer to as the Hall crystal and Wigner crystal for brevity). However, it is straight forward to extend our discussions the the honeycomb crystals as well. In Sec.~\ref{sec:FracExtension} we will show that a honeycomb lattice can be stabilized, if we instead consider fractional filling per magnetic flux. 

\subsubsection{Hall liquid}
The Hall liquid phase, $\rho_Q = 0$, was considered in detail in Ref.~\cite{zhang1992chern}, where it was shown to have a quantized Hall conductance, as expected. Here, we will review these arguments, in preparation for our analysis of the Hall crystal. 

To show that the Hall liquid state does indeed have a quantized Hall conductance, we consider fluctuations around the mean-field ansatz,
\begin{equation}\begin{split}
&\Phi = e^{i \delta \theta} \sqrt{1 + \delta\rho} ,
\\&a_\mu = \delta a_\mu.
\label{eq:LiquidSaddleFluct}\end{split}\end{equation}
where $\delta\theta$ is the phase fluctuation of $\Phi$, $\delta\rho$ is the density fluctuation, and $\delta a_\mu$ denotes the gauge field fluctuation. 
Since we are not considering vortices, we can fix $\Phi$ to be real ($\delta\theta = 0$). After using the equations of motion for $\delta a_t$ to integrate out the density fluctuations, $\delta\rho$, the non-interacting Lagrangian for $\delta \bm{a}$ is,
\begin{equation}\begin{split}
    \mathcal{L} = &-\frac{1}{2} \delta\bm{a} \phantom{|}\Pi_{\bm{p}}\phantom{|}\delta\bm{a}+ \frac{\omega }{4\pi} \left( \delta \bm{a} + \tilde{\bm{A}}\right) \Sigma^y \left(\delta \bm{a} + \tilde{\bm{A}} \right),
\label{eq:fluctuatingLiquidGauge}\end{split}
\end{equation}
where $\Sigma^y$ is the Pauli matrix acting on the spatial indices of the gauge fields, (i.e., $\delta \bm{a}\Sigma^y \delta \bm{a} = -i \delta \bm{a}_y\delta \bm{a}_x + i \delta \bm{a}_x\delta \bm{a}_y$). The couplings between the gauge field and the bosonic matter are encoded in the $2\times 2$ matrix $\Pi$. We have also included additional \textit{probe} gauge fields $\tilde{\bm{A}}$.

After integrating out all remaining fluctuations, we arrive at the following electromagnetic response theory,
\begin{equation}\begin{split}
&\mathcal{L}_{\text{response}} = \tilde{\bm{A}}  \phantom{|} \Pi^{\text{eff}}_{\omega,\bm{p}}\phantom{|} \tilde{\bm{A}},\\
&\Pi^{\text{eff}}_{\omega,\bm{p}} = \frac{\omega^2}{32 \pi^2} \Sigma^y \left[ {\Pi} _{\bm{p}} + \frac{\omega}{4\pi}\Sigma^y \right]^{-1}  \Sigma^y + \frac{\omega}{4\pi} \Sigma^y. 
\end{split}\end{equation}
In terms of $\Pi^{\text{eff}}_{\omega,\bm{p}}$, the Hall conductance is, 
\begin{equation}\begin{split}
\sigma_H = \lim_{\omega\rightarrow 0}\frac{1}{\omega}\text{Tr}\left[\Pi^{\text{eff}}_{\omega,\bm{p}=0}\Sigma^y\right].
\label{eq:LiquidHall}\end{split}\end{equation}
It is straightforward to show that ${\Pi}_{\bm{p} = 0} = \Sigma^0$,
the 2$x$2 identity. Using this, we find a quantized Hall conductance of $\sigma_{H} = \frac{1}{2\pi}$, as expected. In general, $\sigma_{H} = \frac{1}{2\pi}$ as long as ${\Pi}^{-1}_{\bm{p}= 0}$ is non-singular. This is true when the gauge fields $a_\mu$ are Higgsed.

\subsubsection{Hall crystal}
To show that $0 \leq \rho_Q \leq 2/3$ corresponds to a Hall crystal, we must verify that it has Hall conductance $\sigma_H = 1/2\pi$. Since $\Phi$ is non-zero everywhere for $0 \leq \rho_Q \leq 2/3$, the gauge fields are expected to remain Higgsed, and, based on the previous analysis, the Hall conductance will indeed be $\sigma_H = 1/2\pi$. To confirm this, we follow the same steps used above: we consider Gaussian fluctuations around the crystal saddle point defined in Eqs.~\ref{eq:crystalAnsatzDensity} and~\ref{eq:crystalAnsatzGauge}, and then integrate out the fluctuations of $\delta a_t$ and $\delta \rho$ using their equations of motion. This will lead to a Lagrangian for the gauge field fluctuations $\delta \bm{a}$ and $\tilde{\bm{A}}$.

Since the crystal state has discrete translation symmetry, the gauge field fluctuations form bands with well defined crystal momentum, $\bm{k}$. In terms of these bands, the low energy Lagrangian for $\delta \bm{a}$ and $\tilde{\bm{A}}$ is
\begin{equation}\begin{split}
    \mathcal{L} = \sum_n \Big[&-\frac{1}{2}\delta\bm{a}_{n} \phantom{|}\Pi_{\bm{k},n}\phantom{|}\delta\bm{a}_n\\ &+ \frac{\omega }{4\pi} \left( \delta \bm{a}_n + \tilde{\bm{A}}_n \right) \Sigma^y \left(\delta \bm{a}_n + \tilde{\bm{A}}_n \right)\Big]
\label{eq:fluctuatingCrystalGauge}\end{split}
\end{equation}
where $n$ is the band index for the gauge fields. The crystal momentum, $\bm{k}$, occupies the triangular lattice Brillouin zone. The dispersion of the gauge field bands are determined by the $\Pi_{\bm{k},n}$, which is a $2\times 2$ matrix for fixed band index, $n$ and crystal momentum $\bm{k}$. 

After integrating out the remaining emergent gauge field fluctuations, the effective response Lagrangian for the probe gauge fields is,
\begin{equation}\begin{split}
&\mathcal{L}_{\text{response}} = \tilde{\bm{A}}_n  \phantom{|} \Pi^{\text{eff}}_{\omega,\bm{k},n}\phantom{|} \tilde{\bm{A}}_n,\\
&\Pi^{\text{eff}}_{\omega,\bm{k},n} = \frac{\omega^2}{32 \pi^2} \Sigma^y \left[ {\Pi} _{\bm{k},n} + \frac{\omega}{4\pi}\Sigma^y \right]^{-1}  \Sigma^y + \frac{\omega}{4\pi} \Sigma^y. 
\label{eq:CrystalResponse}\end{split}\end{equation}
As long as $\Pi_{\bm{k} = 0,n}$ is invertible, the response theory reduces to the usual Chern Simons term when $\bm{k} = 0$, and $\omega\rightarrow 0$, and the Hall conductance is $\sigma_H = 1/2\pi$. We therefore conclude the the Hall conductance is finite as long as the bands described by $\Pi_{\bm{k},n}$ are gapped.

\begin{figure}
\includegraphics[width = .9\linewidth]{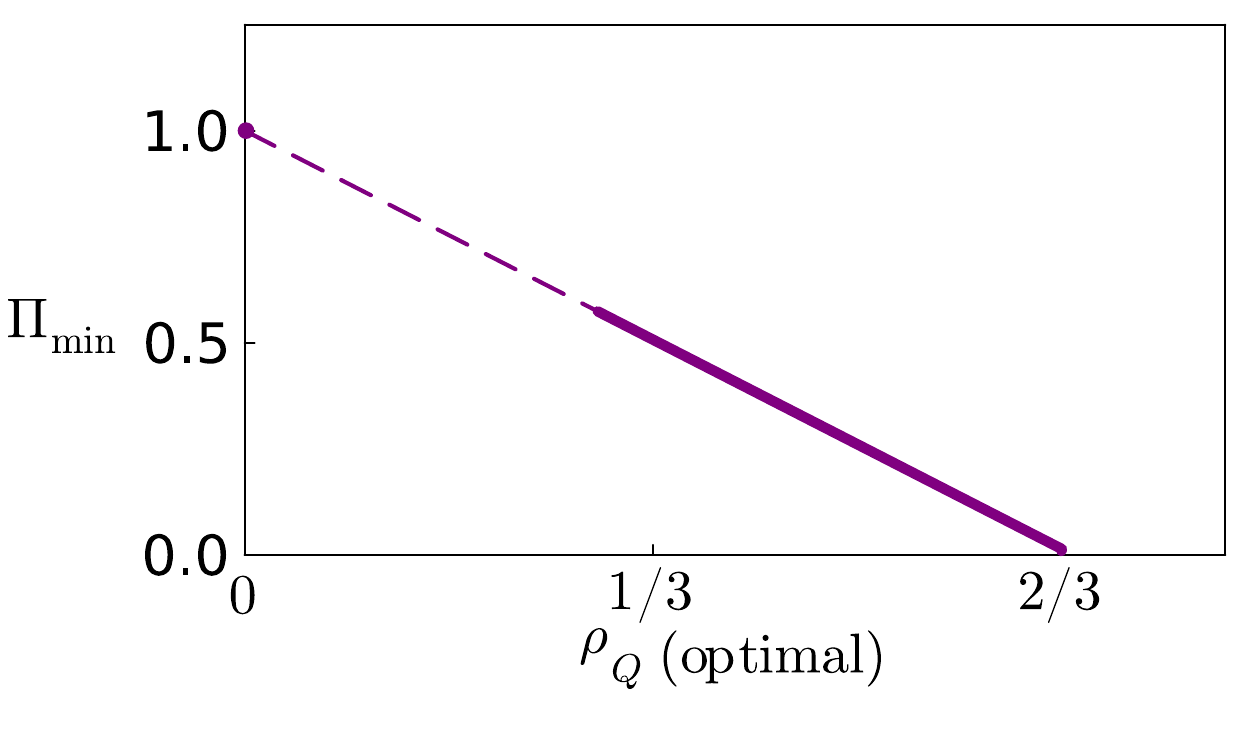}
\caption{The lowest eigenvalue of $\Pi_{\bm{k}=0,n}$ ($\Pi_{\text{min}}$) as a function of the optimal value of $\rho_Q$, calculated using a finite momentum grid with a cutoff of $16|Q^*|$. The dashed line extrapolates between the liquid where $\rho_Q = 0$ is optimal, and the crystal states, where the optimal $\rho_Q$ lies between $\sim .29$ and $2/3$.   }\label{fig:GapsFig}
\end{figure}

In Fig.~\ref{fig:GapsFig} we plot the minimum eigenvalue of $\Pi_{\bm{k} = 0,n}$, $\Pi_{\text{min}}$ as a function of the optimal value of $\rho_Q$. Calculations were done on a finite size momentum grid using $V_C = 50$ and $-45 \omega^2_c <\Delta < 15\omega^2_c$. For each $\Delta$, we find the optimal value of $\rho_Q$ and $\Pi_{\text{min}}$. As we can see, $\Pi_{\text{min}}$ is a linear function of $\rho_Q$, and reaches zero when $\rho_Q = 2/3$ is optimal. As the gauge field fluctuations are massive when the optimal $\rho_Q$ is between $0$ and $2/3$, we confirm that the system is a Hall crystal when the optimal $\rho_Q$ lies in this range.


\subsubsection{Wigner crystal}
\begin{figure}
\includegraphics[width = .9\linewidth]{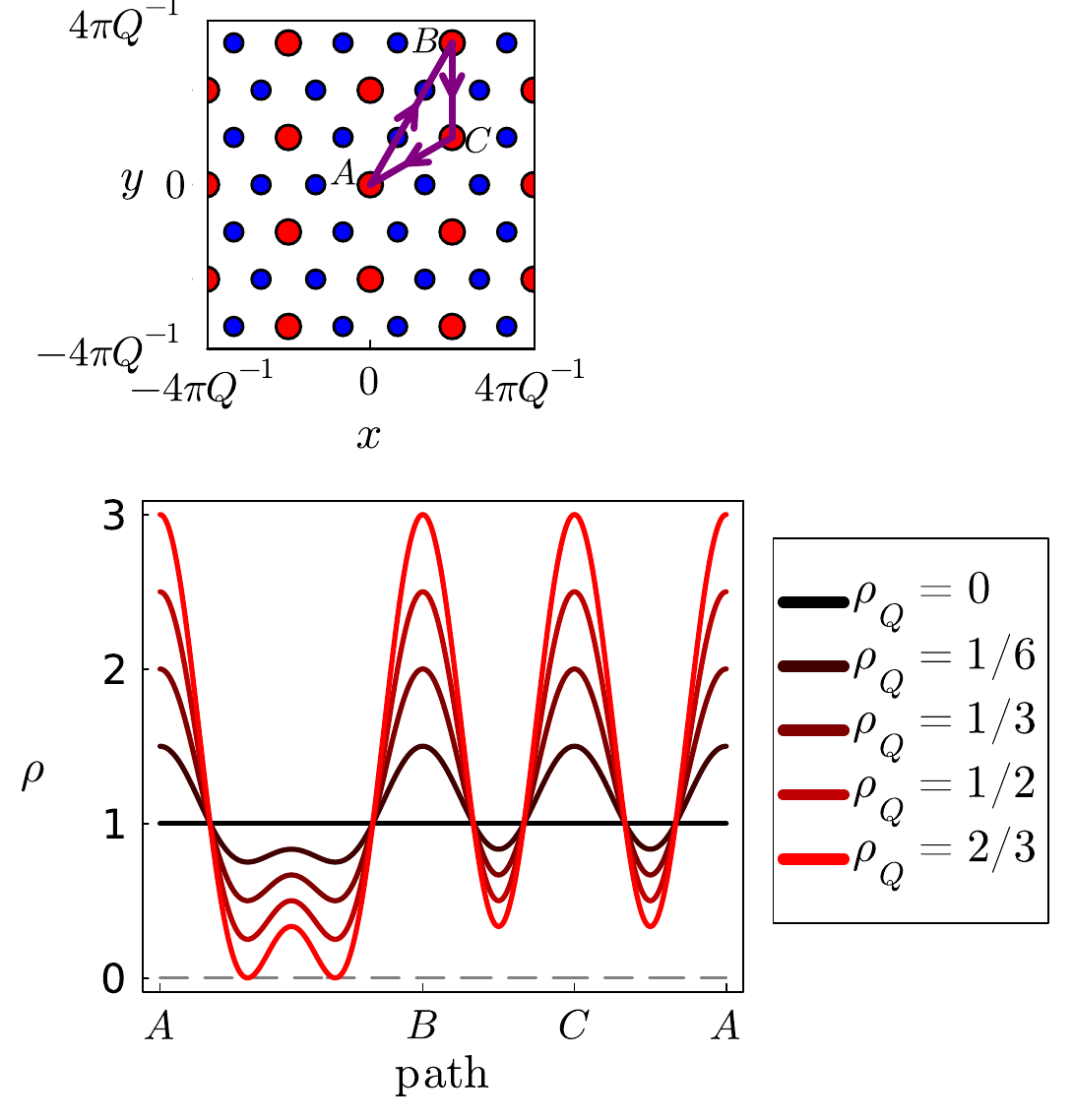}
\caption{Top) location of the maxima (red) and minima (blue) of the crystal ansatz for $\rho_Q > 0$. Bottom) the density profile for different values of $\rho_Q $, for along the real space line cut shown in purple.  }\label{fig:DensityProfileTopoTrans}
\end{figure}

To show that $\rho_{Q} = 2/3$ is a Wigner crystal, we first note that at this value of $\rho_{Q}$, there is a discrete lattice of points where $\Phi = 0$. These zeros occur at the sites of the dual honeycomb lattice, as shown in Fig.~\ref{fig:DensityProfileTopoTrans}. Explicitly, the zeros occur at,
\begin{equation}\begin{split}
\{ \bm{R}_{H} \}= \left\{ n_1 \bm{a}_1 \!+ n_2 \bm{a}_2 + \bm{\tau}_s | n_i \!\in \mathbb{Z},  s\in \!\{A,B\} \right\}
\end{split}\end{equation}
where $\bm{a}_{1,2}$ are the lattice vectors of the trinagular lattice, satisfying $\bm{a}_i \cdot \bm{Q}_i = \delta_{ij}$ for $i=1,2$, and $\bm{\tau}_{A} = \frac{1}{3}(\bm{a}_{1}+\bm{a}_{2})$, $\bm{\tau}_{B} = \frac{2}{3}(\bm{a}_{1}+\bm{a}_{2})$ are the sublattice displacements. 

Because of the zeros of $\Phi$, the gauge symmetry is partially restored. For any function $\Lambda$ that is only non-zero for $\bm{R}\in \{  \bm{R}_H \}$, the system is invariant under the gauge transformation,
\begin{equation}\begin{split}
&\Phi \rightarrow \Phi e^{i\Lambda},\\
&a_\mu \rightarrow a_\mu + \partial_\mu \Lambda.
\label{eq:LatticeGauge_Sym}\end{split}\end{equation}
Clearly, any such $\Lambda$ is highly structured within a given unit cell. However, if we coarse grain and take the long wavelength limit, the short distance fluctuations of $\Lambda$ are smoothed out. The partial gauge symmetry therefore acts like a conventional gauge symmetry in the long wavelength limit. To show this, consider an arbitrary function $\bar{\Lambda}$, that varies slowly over distances $\sim Q^{-1}$. The mean-field Lagrangian invariant is invariant under the gauge transformation, 
\begin{equation}\begin{split}
\Lambda(t,\bm{r}) = \sum_{\bm{R}\in \{  \bm{R}_H \}}\bar{\Lambda}(t,\bm{r}) \delta^2(\bm{r}-\bm{R}),
\end{split}\end{equation}
If we coarse grain $\Lambda$ over distances $\sim Q^{-1}$, it reduces to $\bar{\Lambda}$, as can be confirmed from simple Fourier analysis. The unconventional gauge symmetry of Eq.~\ref{eq:LatticeGauge_Sym} leads to a conventional gauge symmetry in the long wavelength limit. This agrees with the critical theory of Eq.~\ref{eq:MatterLag}, where the Wigner crystal phase is symmetric under gauge transformations.

Due to the zeros of $\Phi$ it is also possible add/remove magnetic fluxes at zero energy cost. Consider a localized flux of the background gauge field at a fixed $\bm{R}\in \{  \bm{R}_H \}$, $\bm{A}\rightarrow \bm{A} +\bm{A}'$ with,
\begin{equation}\begin{split}
\bm{A}'(\bm{r}) = \frac{\hat{z}\times (\bm{r}-\bm{R})}{|\bm{r}-\bm{R}|^2}.
\label{eq:externalFlux}\end{split}\end{equation}
It is possible to compensate for this extra magnetic flux with a bosonic vortex, $\Phi \rightarrow \Phi e^{i\delta \theta}$,
\begin{equation}\begin{split}
&\theta'(\bm{r}) = \tan^{-1}\left(\frac{r_y - R_y}{r_x - R_x}\right).
\label{eq:vortexSolution}\end{split}\end{equation}
A straightforward calculation shows that Eqs.~\ref{eq:externalFlux} and~\ref{eq:vortexSolution} satisfy the flux attachment constraint, and leave the mean-field Hamiltonian invariant when $\rho_{Q} =  2/3$. This insensitivity to an external magnetic field shows that the composite bosons are a Wigner crystal with $\rho_Q = 2/3$. Here, we also explicitly see that magnetic fluxes act as a source for vortices of $\Phi$, same as in the vortex theory in Eq.~\ref{eq:VortexLag}.

It is worth remarking that arguments used here apply to any crystal configuration with a discrete lattice of points where $\rho = 0$. The fact that the Wigner crystal occurs at the maximal allowed value of $\rho_Q$ is an artifact of only considering the first harmonic of the triangular lattice crystal. If one includes higher harmonics of the crystal, the Wigner crystal transition can generically occur before $\rho_Q$ reaches its maximal value.

\section{Extension to Fractional Quantum Hall States}\label{sec:FracExtension}

\begin{figure}[t]
\includegraphics[width = .9\linewidth]{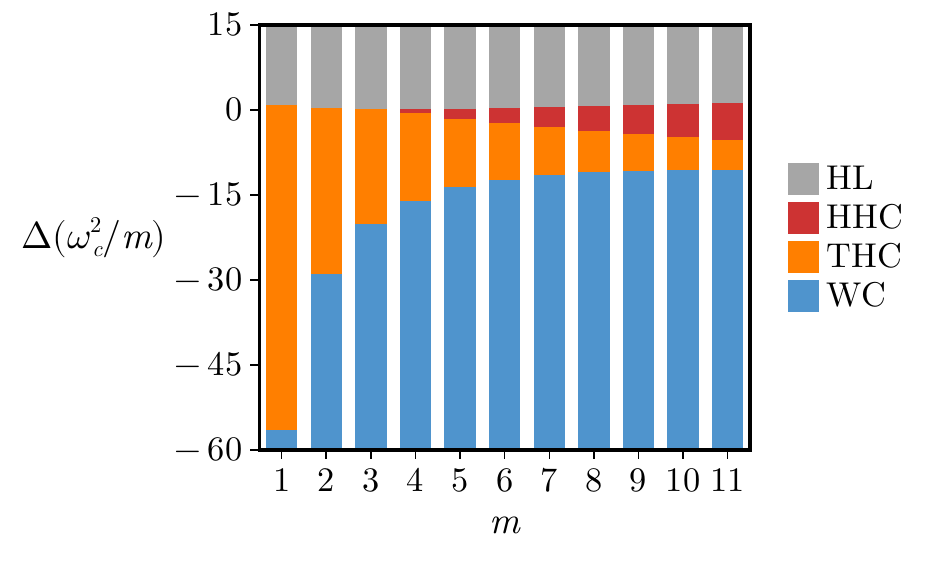}
\caption{Phase diagram as a function of roton mass $\Delta$, and the fluxes bound to each boson $m$. HL is the Hall liquid phase, HHC is the honeycomb Hall crystal, THC is the triangular Hall crystal, WC is the triangular Wigner crystal. 
}\label{fig:frac_phase_diagram}
\end{figure}

It is straightforward to extend the mean field crystals considered here to fractional filling $\nu = 1/m$. At the level of the crystal ansatz, this amounts to rescaling Eq.~\ref{eq:crystalAnsatzGauge} by $m$ to satisfy the flux attachment conditions,
\begin{equation}\begin{split}
\bm{\nabla}\times \bm{a} &= 2\pi m \left( \rho - 1 \right). 
\label{eq:fluxAttachEqsFrac}\end{split}\end{equation} 
The new ansatz now describes a fractional Hall liquid, a fractional Hall crystal---a state with broken translation symmetry and a fractional Hall conductance~\cite{tan2025variational}---and a Wigner crystal. All other aspects of our analysis can be done as before. For electronic systems, $m$ must be odd, but for completeness we will consider both even and odd $m$ here. 

Before discussing the results, it will be useful to examine the assumptions implicit in the density ansatz in Eq.~\ref{eq:crystalAnsatzDensity}. As discussed previously, our ansatz assumes that there are no vortices in the crystal. Vortices of $\Phi$ are anyons at fractional filling, and are the low energy excitation of the liquid state. As anyons are not present in the Wigner crystal, our ansatz is most relevant when considering approaching the fractional Hall crystal as a descendant of a Wigner crystal, rather than viewing it as a descendant of the fractional Hall liquid. The possibility of an ``anyon crystal", which would be a vortex crystal of composite bosons is an interesting possibility, which we leave to future work.

The phase diagram for different $m$ as a function of the roton mass, is given in Fig~\ref{fig:frac_phase_diagram}. Note that for fractional fillings the extrapolated roton mass is 
\begin{equation}\begin{split}
\Delta = 4\pi^2 m^2 + \frac{Q^{4}}{4} + 2\pi V_C Q  + 2\pi  V_R d Q   J_1(Q d). 
\end{split}\end{equation} 
For $1 \leq m \leq 3$ we only find triangular lattice crystals, $\rho_Q > 0$. However, for $m \geq 4$, we find honeycomb lattice crystals where $0 > \rho_Q > -1/3$. Following our earlier discussion, this describes a honeycomb Hall crystal. For the parameters considered here, we do not find any regime where $\rho_Q = -1/3$, (which describes a honeycomb crystal with vanishing Hall conductance), although such a state can occur in principle. 

For $ m \geq 4$, the honeycomb Hall crystal phase occurs between the Hall liquid and the triangular lattice Hall crystal. The liquid to honeycomb Hall crystal transition is first order, for the same reasons as before. The Hall crystal to Hall crystal transition is also first order as $\rho_Q$ abruptly changes sign.

In general, we find triangular crystals are preferred for strong interactions (large $\Delta$) while the honeycomb Hall crystal appears at large $m$ and intermediate interaction strength. This behavior can be understood in terms of a competition between interaction and kinetic energies of the bosons. Since the interactions between the bosons are repulsive, the interaction energy is minimized when the density maxima are maximally far apart from each other. This favors triangular crystals. For low filling (large $m$), the dominant contribution to the kinetic energy is from the diamagnetic term, $\frac{1}{2}|\bm{a}|^2\rho \propto m^2$. Since $|\bm{a}|^2 \propto (\rho-1)^2$, the energetic penalty for positive density fluctuations ($\rho-1 > 0$) is larger than for negative density fluctuations ($\rho-1 < 0 $). An asymmetry of this form is expected as the $1/m$ fractional quantum Hall states are strongly particle-hole asymmetric, with quasi-holes being lighter than quasi-particles~\cite{laughlin1983anomalous, haldane1983fractional}. In the honeycomb crystal, positive density fluctuations are smeared out over a large region than they are in the triangular crystal, which reduces the energetic cost from the diamagnetic term. This leads to honeycomb crystals being  energetically favored at large $m$, and intermediate interaction strength.

We conjecture that honeycomb Hall crystals will also be preferred at large $m$ if one considers a crystal with vortices, i.e. an anyon crystal. If one were to form a six-fold symmetric crystal out of these vortices, and keep the system at rational filling, the minimal options are to have a $\pm 2$-winding vortex at the site of the triangular lattice and a $\mp 1$-winding vortex at each site of the dual honeycomb lattice. In Ref.~\cite{tafelmayer1993topological}, it was shown that positive winding vortices repel each more strongly than negative winding vortices in composite boson theory. This indicates that the crystal with $-2$ vortices on the triangular lattice sites and $+ 1$ vortices on the honeycomb lattices sites will have lower energy. 
Physically, this will be a honeycomb lattice of charge $1/m$ anyons, on a triangular background of charge $-2/m$ anyons. However, if the repulsive interactions are significantly strong and long ranged, the opposite configuration may be preferred, as it maximizes the distances between positive charges. Verifying this conjecture is left to future work. 

As a final note, we point out that for the fractional cases, the transition between the Hall crystal and Wigner crystal is no longer described by a Dirac fermion. The matter theory can instead be written in terms of a Dirac fermion coupled to two emergent gauge fields, $b$ and $c$,\cite{lee2018emergent}
\begin{equation}\begin{split}
\mathcal{L}_{\text{mat}} =\: &i \bar{\Psi} \slashed{D}^{c} \Psi - m_D \bar{\Psi} \Psi \\ & + \frac{1}{8\pi} cdc - \frac{m-1}{4\pi} bdb + \frac{1}{2\pi} bd(c-\tilde{A}) .
\label{eq:DiracDualized2}\end{split}\end{equation}
Due to the strong coupling between the gauge fields and $\Psi$ it is no longer possible to use power counting arguments to conclude that the phonon coupling is irrelevant at this transition. The effects of phonons on Eq.~\ref{eq:DiracDualized2}, is an interesting problem for future work.

\section{Conclusion}\label{sec:conclusion}
In this work, we have analyzed the possibilities of Hall liquids, Hall crystals, and Wigner crystals and the transitions between them using composite boson theory. We found a first order transition between the liquid and Hall crystal phases, and a continuous topological transition between the Hall and Wigner crystals. At the critical point between the crystalline phases, phonons are irrelevant, and the transition is described by a single Dirac fermion. 

From our mean-field analysis, we found that a soft roton mode can induce a Hall crystal state. For integer filling, $\nu = 1$, the Hall crystal is triangular. However, for low fractional fillings, $\nu = 1/m$, and intermediate interaction strength, it is possible to have honeycomb Hall crystals. As the extrapolated roton mass continues to decrease, the composite boson vortices proliferate, and the Hall crystal transitions into a Wigner crystal. 

In a real system, a Hall crystal would display both a quantized Hall resistance, as well as display a periodic density modulation. The latter can be visualized with a local probe, like scanning tunneling microscopy (STM). In Ref. \cite{tsui2024direct} Wigner crystals were imaged using STM on a 2DEG in a magnetic field. Combined with transport, this procedure could be used to verify the existence of a Hall crystal, and, potentially, the honeycomb crystals we predicted to occur at fractional fillings. 

There are several directions for future work. First, there is the extension of our analysis to anomalous partial Hall crystals, which are believed to underlay the anomalous Hall state observed in rhombohedral graphene~\cite{dong2024anomalous}. It is also possible to consider systems in a magnetic field that are doped away from rational filling. In composite boson theory, the superconductor would experience a net magnetic field. If one is near fractional filling, the resulting Abrikosov vortex lattice~\cite{abrikosov1957magnetic} would be an anyonic Hall crystal, described by both a rational $\nu$ and $\eta$ in Eq.~\ref{eq:generalizedStreda}. Finally, the effects of phonons at the fractional Hall crystal to Wigner crystal transition need to be fully analyzed. 

\section*{Acknowledgments}
We thank J Dong, I Esterlis, H Goldman, S Kivelson, P Nosov, and D Shi for helpful discussions. JMM is supported by a Leinweber Institute for Theoretical Physics fellowship. SB and SR are supported in part by the US Department of Energy, Office of Basic Energy Sciences, Division
of Materials Sciences and Engineering, under Contract
No. DE-AC02-76SF00515S.

\bibliography{CB_HC.bib}
\bibliographystyle{apsrev4-1}
\appendix

\widetext

\section{Landau theory for triangular lattice crystallization}\label{app:FreeEnergyTrianglular}
For a triangular lattice crystal, we have three order parameters $\rho_{\bm{Q}_i}$ $i = 1,2,3$. These measure the density modulations at $\bm{Q}_i$, the reciprocal lattice vectors of the crystal. Expanding the free energy is near the liquid crystal transition,
\begin{equation}
\begin{split}
    \mathcal{F} = & \sum_i r |\rho_{\bm{Q}_i}|^2  + w \rho_{\bm{Q}_1}\rho_{\bm{Q}_2}\rho_{\bm{Q}_3} + \sum_i u |\rho_{\bm{Q}_i}|^4 + \sum_{i<j} u' |\rho_{\bm{Q}_i}|^2 |\rho_{\bm{Q}_i}|^2 
\end{split}
\end{equation}
where $r$ is the tuning knob, which, in the liquid state, is the mass of density fluctuations with wavenumber $|\bm{Q}_i|$. The tri-linear term $w$, is allowed by symmetry, as $\bm{Q}_1+\bm{Q}_2+\bm{Q}_3= 0$ for the triangular lattice. Because of this term, the transition is first order and occurs at $r> 0$. This term is not present for other translation symmetry breaking orders (e.g., square lattice crystals, or striped phases) and the transition is instead continuous and occurs at $r = 0$ within Landau theory. This term also lowers the free energy of the triangular lattice crystal compared to other translation symmetry breaking orders. 

\section{Boson-vortex duality}\label{vortex_theory}
The discussion here broadly follows that in Ref.~\cite{zhang1992chern}. We start from the composite boson Lagrangian specified in Eq.~\ref{eq:GenLagrangian}. We first write the boson field in density--phase variables $\Phi = \sqrt{\rho}\,e^{i\theta}$, and substituting into the boson Lagrangian we find,
\begin{equation}
\mathcal{L}_{\text{B}}
=
\rho(\partial_t\theta + a_t - A_t)
-\frac{\rho}{2M^*}
(\bm{\nabla}\theta + \bm{a}-\bm{A})^2
+ \mathcal{L}_{\text{int}}(\rho).
\end{equation}
Next, we introduce the boson current $j^\mu=(\rho,\bm{j})$ using a Hubbard-Stratonovich field, and we obtain,
\begin{equation}
\mathcal{L}_{\text{B}}
=
j^\mu(\partial_\mu\theta + a_\mu - A_\mu)
+\frac{M^*}{2\rho}\bm{j}^2
+\mathcal{L}_{\textrm{int}}(\rho).
\end{equation}
We next split the phase into smooth and vortex parts,
\begin{equation}
\theta = \theta_{\text{sm}} + \theta_v .
\end{equation}
Integrating over $\theta_{\text{sm}}$ imposes a current conservation,
\begin{equation}
\partial_\mu j^\mu = 0 .
\end{equation}
In $2+1$ dimensions, such a conserved current can be written as the curl of a dual gauge field $\alpha_\mu$, that is,
\begin{equation}
j^\mu
=
\frac{1}{2\pi}
\epsilon^{\mu\nu\lambda}\partial_\nu \alpha_\lambda.
\end{equation}
One can also define a vortex current given by,
\begin{equation}
j_v^\mu
=
\frac{1}{2\pi}
\epsilon^{\mu\nu\lambda}
\partial_\nu\partial_\lambda \theta_v .
\end{equation}
Substituting these currents back into the action yields,
\begin{equation}
\mathcal{L}_{\text{B}}
=
\frac{1}{2\pi}
\epsilon^{\mu\nu\lambda}
\alpha_\mu \partial_\nu (a_\lambda - A_\lambda)
+
\alpha_\mu j_v^\mu
+ \ldots
\end{equation}
We now write the full Lagrangian. We include the Chern-Simons term so that,
\begin{equation}
\mathcal{L}
= \frac{1}{2\pi}\alpha\,d(a-A)
+
\frac{1}{4\pi m}a\,da
+
\alpha_\mu j_v^\mu
+ \ldots
\end{equation}
We now promote the vortex current to a dynamical vortex field $\varphi_v$, and because of the argument given in the main text, we obtain a relativistic dual Lagrangian given by,
\begin{equation}\begin{split}
\mathcal{L}_{\textrm{vort}} = \frac{1}{2} |D^\alpha_\mu \phi_v|^2 - r_v|\phi_v|^2 - u |\phi_v|^4 - \frac{1}{2\pi} \alpha d (a-A) + \frac{1}{4\pi} ada, 
\label{eq:VortexLag}\end{split}\end{equation}
where  $D^\alpha_\mu \equiv \partial_\mu - i \alpha $, and  $ada \equiv \epsilon^{\mu\nu\lambda}a_\mu\partial_\nu a_\lambda$. The coarse grained vortex fluctuations are encoded in $\phi_v$, and the new gauge field $\alpha$ enforces that vortices of $\Phi$ are accompanied by fluxes of $a-A$, (see Eq.~\ref{eq:GenLagrangian}). Integrating out the gauge field $a$, one obtains the vortex Lagrangian in Eq.~\ref{eq:VortexLag}.

\end{document}